\documentclass[10pt,journal,letterpaper,compsoc]{IEEEtran}
\usepackage{amsmath}

\usepackage{epsfig,subfigure}
\usepackage[ruled,vlined]{algorithm2e}

\providecommand{\eat}[1]{}

\begin{document}
\title{{Case For Static AMSDU Aggregation in WLANs}}
\author{\authorblockN{Gautam Bhanage\authorrefmark{2}} \\
\authorblockA{\authorrefmark{2}Tech Report: Bhanage.com / GDB2017-006} \\
gbhanage@gmail.com  \\
July 2017}

\maketitle

\begin{abstract}
Frame aggregation is a mechanism by which multiple frames are combined into a single transmission unit over the air. Frames aggregated at the AMSDU level use a common CRC check to enforce integrity. For longer aggregated AMSDU frames, the packet error rate increases significantly for the same bit error rate.  Hence, multiple studies have proposed doing AMSDU aggregation adaptively based on the error rate. This study evaluates if there is a \emph{practical} advantage in doing adaptive AMSDU aggregation based on the link bit error rate. Evaluations on a model show that instead of implementing a complex adaptive AMSDU frame aggregation mechanism which impact queuing and other implementation aspects, it is easier to influence packet error rate with traditional mechanisms while keeping the AMSDU aggregation logic simple.
\end{abstract}

%
\IEEEpeerreviewmaketitle

\section{Introduction}
\PARstart{F}rame aggregation is supported in the WiFi 802.11~\cite{80211n} standard at two levels~\cite{bhanage2017amsdu}. AMSDU (aggregate MAC service data unit) level aggregation operates at the MAC service data unit layer and allows for the reduction of frame size while carrying a single CRC check. As aggregation packs more number of bytes together, it improves MAC efficiency by lowering the airtime required for transmissions. It has been shown in the past that aggregation is a basis for all other MAC optimizations like virtualization~\cite{Bhanage10Virtual, bhanage2017mixing} or multicast optimization~\cite{bhanage2016enhanced} to work well.

\subsection{Problem Statement}
So if sending larger aggregates is better why not always aggregate? The tutorial~\cite{bhanage2017amsdu} clearly explains the tradeoffs between doing aggregation at different layers and the amount of aggregation that is optimal. One of the main concerns is that, when aggregating at the AMSDU layer, this results in us having a single CRC check for the frame, i.e. if the frame fails due to signal strength or collisions at the receiver, then the entire aggregate needs to be retried. In this case, should we retry the payload of the aggregate as a single transmission unit or should it be retried as smaller units to bring down the packet error rate (PER). Or is it better to leave the aggregate as is, and instead change some other MAC layer parameter like the transmit rate?

\begin{figure}
	\begin{center}
		\includegraphics[scale=0.55,angle=0]{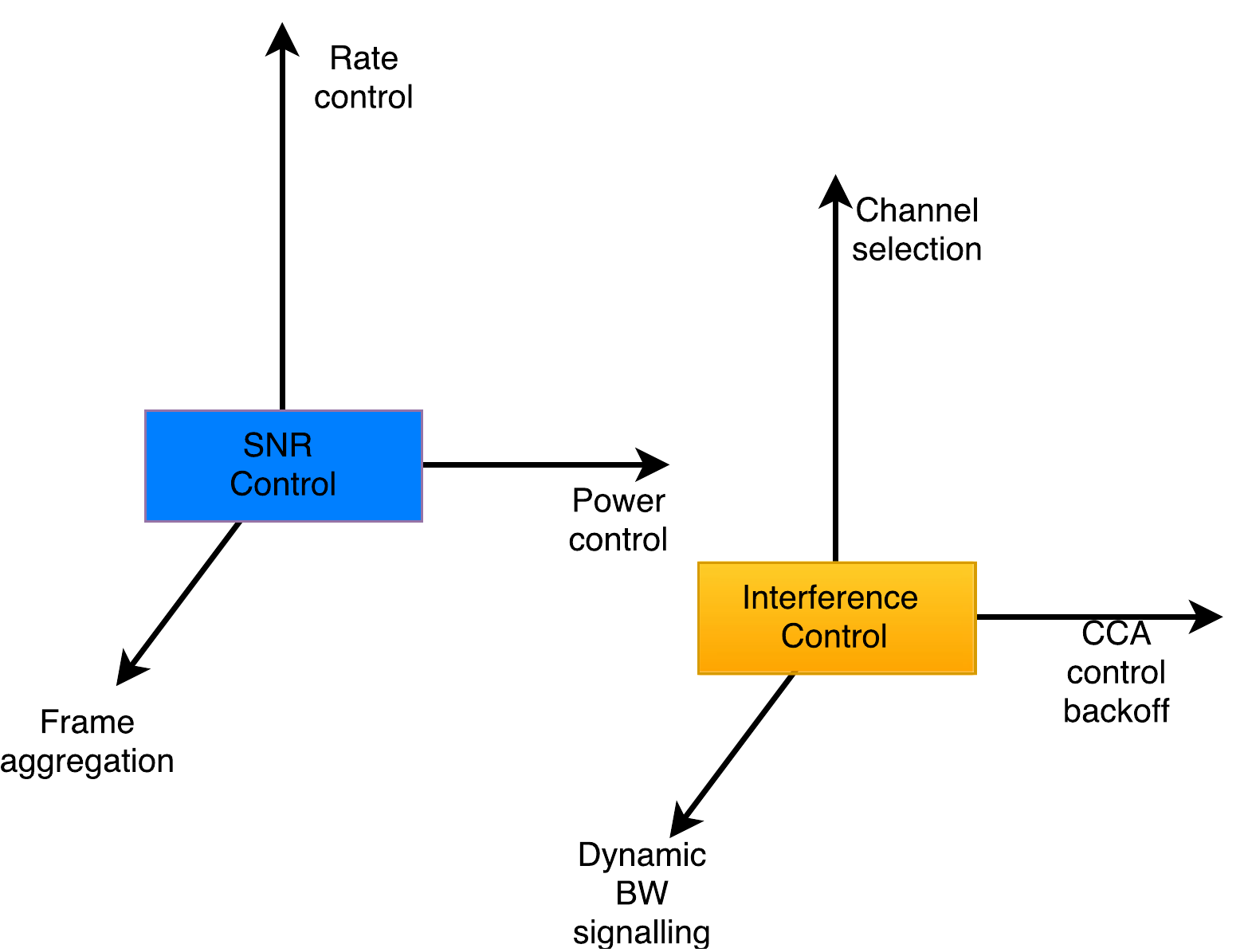}
		\caption{Different dimensions of controls on 802.11 transmissions that can be used for achieving different packet error rates. For controlling SNR, we can use mechanisms such as rate control, frame aggregation control or power control. For controlling the interference seen on a link we can use mechanisms like channel selection, bandwidth selection or CCA control.} \label{fig:dimemsions}
	\end{center}
\end{figure}

Figure~\ref{fig:dimemsions} shows the some of the aspects that can be controlled on a per packet basis to achieve the desired PER\footnote{This list of mechanism is just an example and is in no way comprehensive.}. As shown, losses due to interference can be controlled with a number of mechanisms like: (1) channel selection, (2) CCA mechanisms in CSMA, or (3) dynamic bandwidth signaling and other mechanisms. Any or all of these mechanisms can be applied on a broad scale for interference reduction, and thus limit PER. Losses due to a low SNR at the receiver can be controlled with (1) power control, (2) frame aggregation control and (3) rate control. As we will explain in Section~\ref{sec:model}, doing adaptive AMSDU aggregation based on PER is not optimal from a practical standpoint. Instead, we examine if similar PER results can be achieved by using rate control while not losing efficiency due to aggregation. 

\subsection{Contributions}
This study will focus on the following aspects of MAC layer aggregation and transmissions:
\begin{itemize}
\item Compare if and why PER should be an important factor to consider while doing AMSDU frame aggregation.
\item Discuss practical aspects of adaptive AMSDU aggregation based on PER.
\item Discuss a model for calculations of PER and airtime.
\item Show through results of 802.11 based modeling that instead of using adaptive AMSDU aggregation the desired PER can be easily achieved by using rate control or other orthogonal mechanisms while not sacrificing on efficiency.
\end{itemize}
The rest of the paper is setup as follows. Section~\ref{sec:related} provides a brief background on the work done in this area. Section~\ref{sec:model} discusses the different aspects of the model we use. Section~\ref{sec:simulations} provides results from our model and the inferences that can be drawn from them. Section~\ref{sec:conc} provides concluding remarks and future direction.

\section{Related Work}
\label{sec:related}
Aggregation based on backlogged traffic is generally useful for improving WLAN efficiency~\cite{bhanage2010using, bhanage2011backlogged, bhanage2016apparatus}. The WiFi standard supports aggregation in two ways. A quick primer~\cite{bhanage2017amsdu} on the MSDU and MPDU level aggregation shows the trade off of implementing aggregation at different layers. A survey of aggregation mechanism in WLANs is discussed here~\cite{surveyagg}. We build on these studies.

\section{Analytical Model}
\label{sec:model}


\subsection{Challenges in Adaptive AMSDU Aggregation}
\label{sec:challenges}
There are a number of factors which decide the efficacy of AMSDU aggregation which is dependent on PER. We consider two cases. (1) Basic transmissions based on PER, and (2) Re-transmissions based on PER.

In case of initial individual transmissions, the transmit mechanism on the software will have to track the error rate seen for different AMSDU lengths, per tx-rate, per tx-power, per peer. With faster standards like 802.11ax, the number of combinations that need to be maintained would be significantly large.  Also, because of reasons like mobility and changing RF environment, this information grows stale over time, which would end up requiring us to do re-transmissions.

Re-transmissions based on PER are extremely inefficient, if possible at all because:
\begin{itemize}
	\item The hardware needs to indicate to the software that the AMSDU aggregate transmission failed. Retries cannot be implemented in hardware.
	\item In the retry case, the frame needs to be either DMA\footnote{Direct memory access is a common memory access and transfer mechanism used between radio hardware and software.} transfered back to the radio software for de-aggregation and then re-aggregation needs to be done with the correct length.
	\item Changing the frame length will also change how queuing is achieved  and subsequent or already transmitted frames are handled. This is to ensure that there is no out of order delivery at the receiver.
	\item Most WLANs will employ some form of security, e.g. WPA or WPA2. In this case, when we are re-aggregating retries we may have to decrypt and re-encrypt the frames in line. This is also a fairly costly proposition.
	\item Apart from the computation costs\footnote{Higher required computation speed is one way to look at the increased cost in the system. If we model the WiFi transmission and service mechanism as an M/G/1 system, the additional arrival rate due to retransmissions would result in a higher arrival rate of $\lambda2$. This would potentially require us to accommodate $\frac{(\lambda2)^2 \sigma  + (\lambda2)/ \mu}{2(1 - (\lambda2)/ \mu)}  -  \frac{(\lambda)^2 \sigma  + (\lambda)/ \mu}{2(1 - (\lambda)/ \mu)}$ more packets.}, depending on the traffic conditions for the software queue, these overheads may also lead to variable delays which will further deteriorate the perceived \emph{user experience} on the link.
	\item All of the above do not take into account the cost of cache misses and DMA transferring other miscellaneous data required for the frame transmission.
\end{itemize}


Based on the above comments, we can see that a practical implementation of adaptive AMSDU aggregation based on PER is fairly complicated. Per packet transmit rate control, on the other hand, is relatively easy to achieve because, some both re-transmissions in hardware at lower rates are possible or re-queuing the same frame to hardware at a lower rate (from software) is easy. The goal of this study is to evaluate if simple \textbf{static} AMSDU aggregation\footnote{For the purpose of this study we consider adaptive AMSDU aggregation only, because the AMSDU is the unit at which retries are implemented. For example, in a case where AMSDUs are included in an AMPDU, individual failing MPDUs or AMSDUs are retried.} can be used in combination with rate control to achieve the desired PER while not sacrificing MAC efficiency.

\subsection{Relation Between PER and Frame Size}
PER is related to frame size in two ways. Frame transmission can fail at the receiver because of (1) low signal to noise ratio (SNR) or it can fail due to (2) interference. In this context, we define the \emph{length} of a packet as the airtime required for transmitting it. In both the cases (noise and intereference), a higher airtime for transmitting the failure will result in higher chances of failure.

In the case of failures due to noise at the receiver, decoding a longer packet at a lower SNR is fairly is likely to see more failures than decoding a smaller packet at the same SNR value:
\begin{equation}
PER = 1 - (1 - BER)^{len}
\end{equation}
Thus as frame length (aggregate length) increases, PER increases exponentially for the same BER.

Losses due to collisions are dependent on the airtime used for frame transmissions, which in turn are directly dependent on the length of the aggregate AMSDU. For the same bit error rate (BER), the packet error rate for an aggregated frame (longer frame with a single CRC) is higher for the aggregate than otherwise.  

Based on the above, we see that in both cases, with interference or noise, longer frames suffer with a higher PER. However, as is shown previously~\cite{bhanage2017amsdu}, if we do not aggregate enough, we lose efficiency.

\begin{figure}[t]
	\begin{center}
		\epsfig{figure=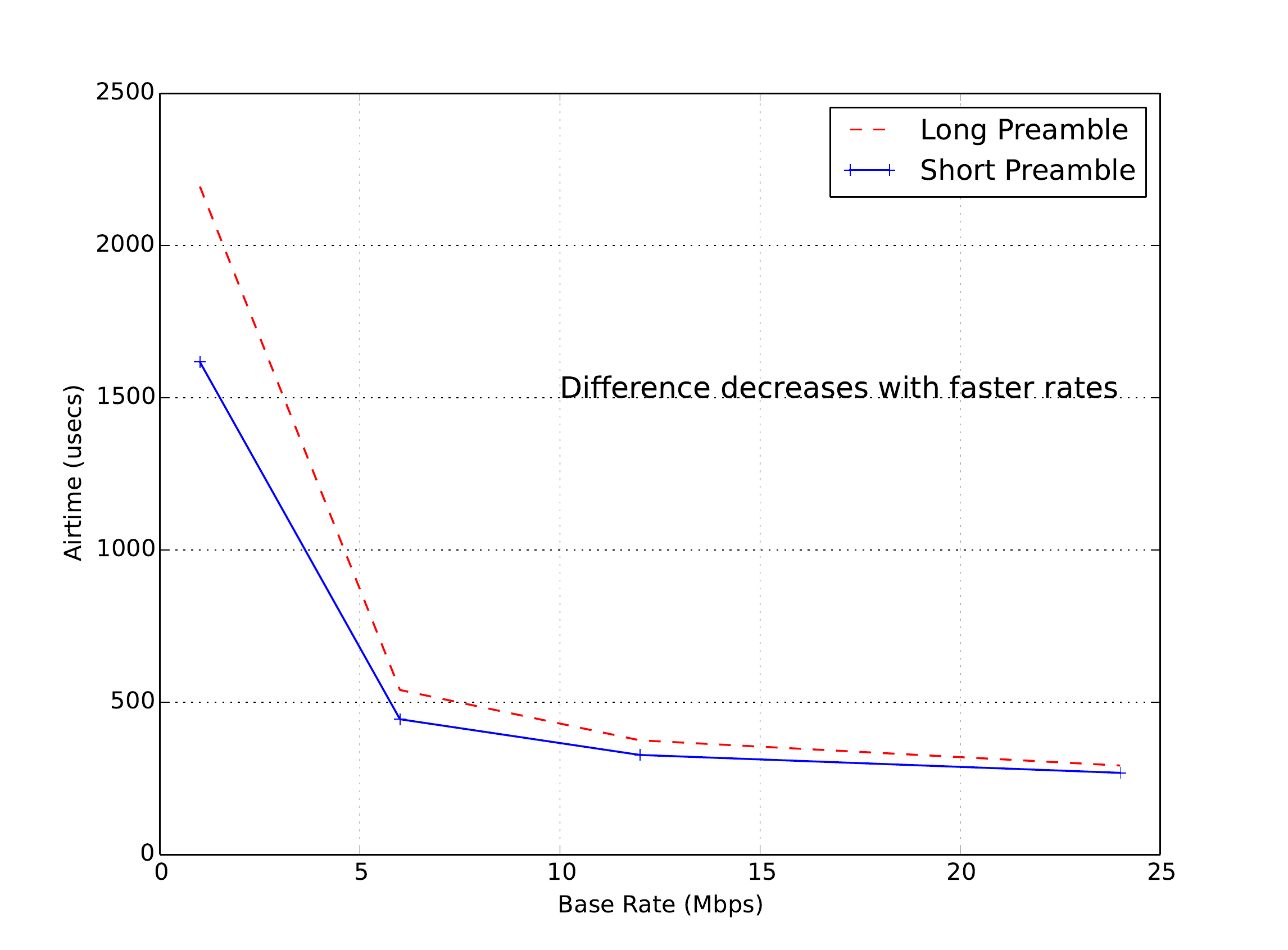,width=2.9in}
		\caption{Weights (airtime allocations for slices) calculated by the HWV controller based on varying load conditions.} \label{fig:basic_rate_airtime}
	\end{center}
\end{figure}

\subsection{Airtime Calculations}
We calculate the airtime required for a regular transmission as follows:

\begin{equation}
Airtime_{ovh1} = Backoff \times T_{diffs}
\end{equation}

\begin{equation}
Airtime_{ovh2} = T_{phy} + T_{mac} + T_{llc} + T_{ack} + T_{sifs} 
\end{equation}

\begin{equation}
Airtime_{overheads} = Airtime_{ovh1} + Airtime_{ovh2}
\end{equation}

\begin{equation}
Airtime_{payload} = \frac{payload size}{physical layer rate}
\end{equation}

Thus, we can now calculate total airtime as: $Airtime_{total} = Airtime_{payload} + Airtime_{overhead}$. As shown in Figure~\ref{fig:basic_rate_airtime}, negotiating a higher basic rate for the WLAN results in a further reduced airtime consumption.

\begin{table}
	\begin{center}
		\caption{Simulation Parameters}
		\begin{tabular}{ l | l  }
			\hline
			Parameter & Value \\ \hline \hline
			$T_{sifs}$ & 10 usecs \\ \hline
			$T_{difs}$ & 50 usecs \\ \hline
			$T_{ack}$ & 14 bytes \\ \hline
			$T_{phy}$ header short & 120 bytes  \\ \hline
			$T_{phy}$ header long & 192 bytes  \\ \hline
			$T_{mac}$ header & 34 bytes \\ \hline
			Data rates & VHT  \\
			\hline
		\end{tabular}
		\label{tab:simulation_params}
	\end{center}
\end{table}

\section{Simulations Results}
\label{sec:simulations}
As discussed in Section~\ref{sec:challenges}, implementing adaptive AMSDU scheme has practical implementation challenges. The goal of this set of evaluations is to ensure that the static AMSDU scheme along with rate control performs as well as a scheme with dynamic AMSDU scheme. We achieve this by validating that our metrics: airtime and PER have similar performance in the case of (1) rate control with static AMSDU and (2) Dynamic AMSDU (the value of which is represented by a different AMSDU size at the same rate). Parameters used for the simulations are as listed in the Table~\ref{tab:simulation_params}.

\begin{figure*}
	\centering
	\begin{subfigure}
		\centering
		\epsfig{figure=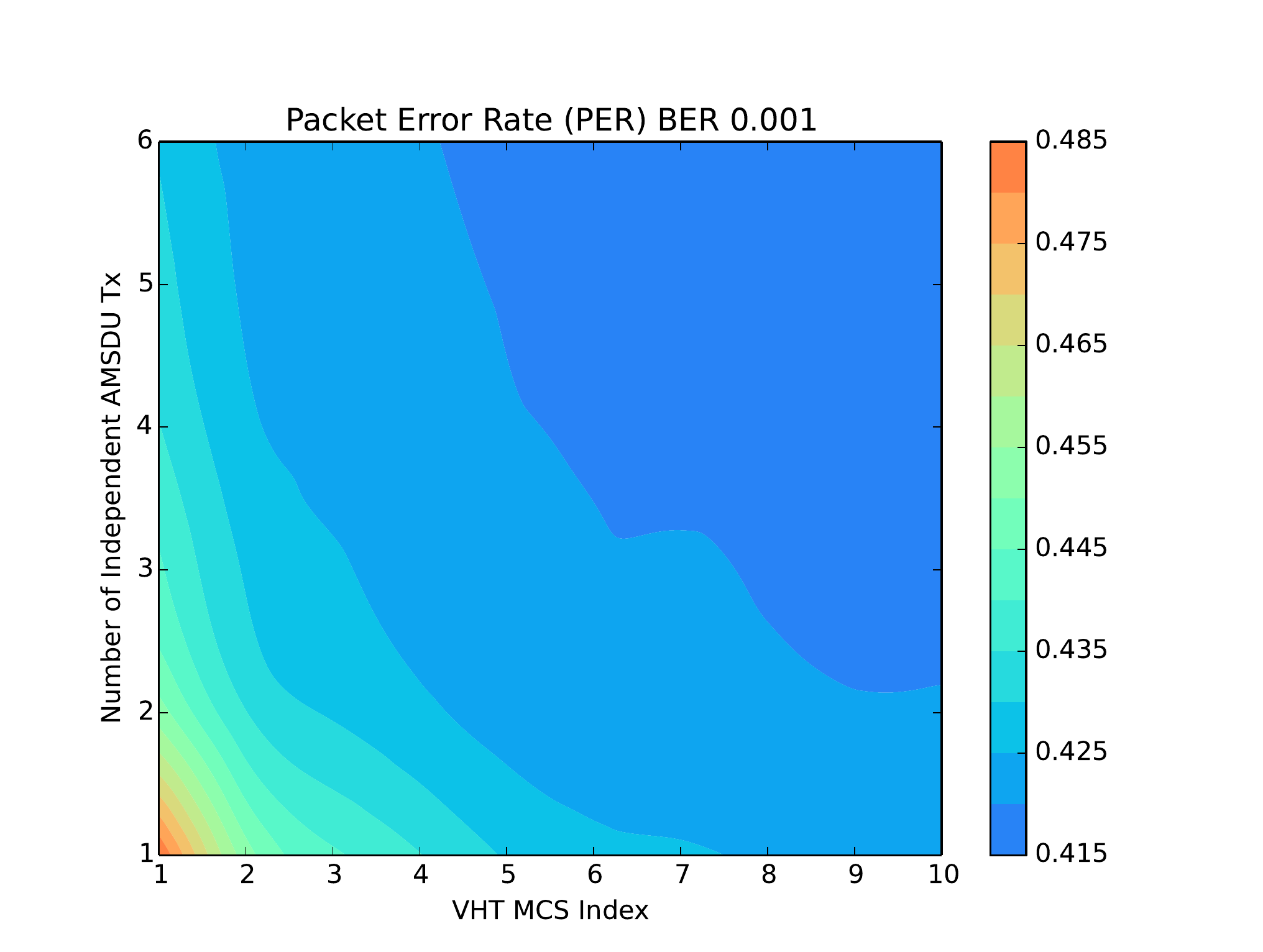,width=2.32in}
		\label{fig:ber001}
	\end{subfigure}
	\begin{subfigure}
		\centering
		\epsfig{figure=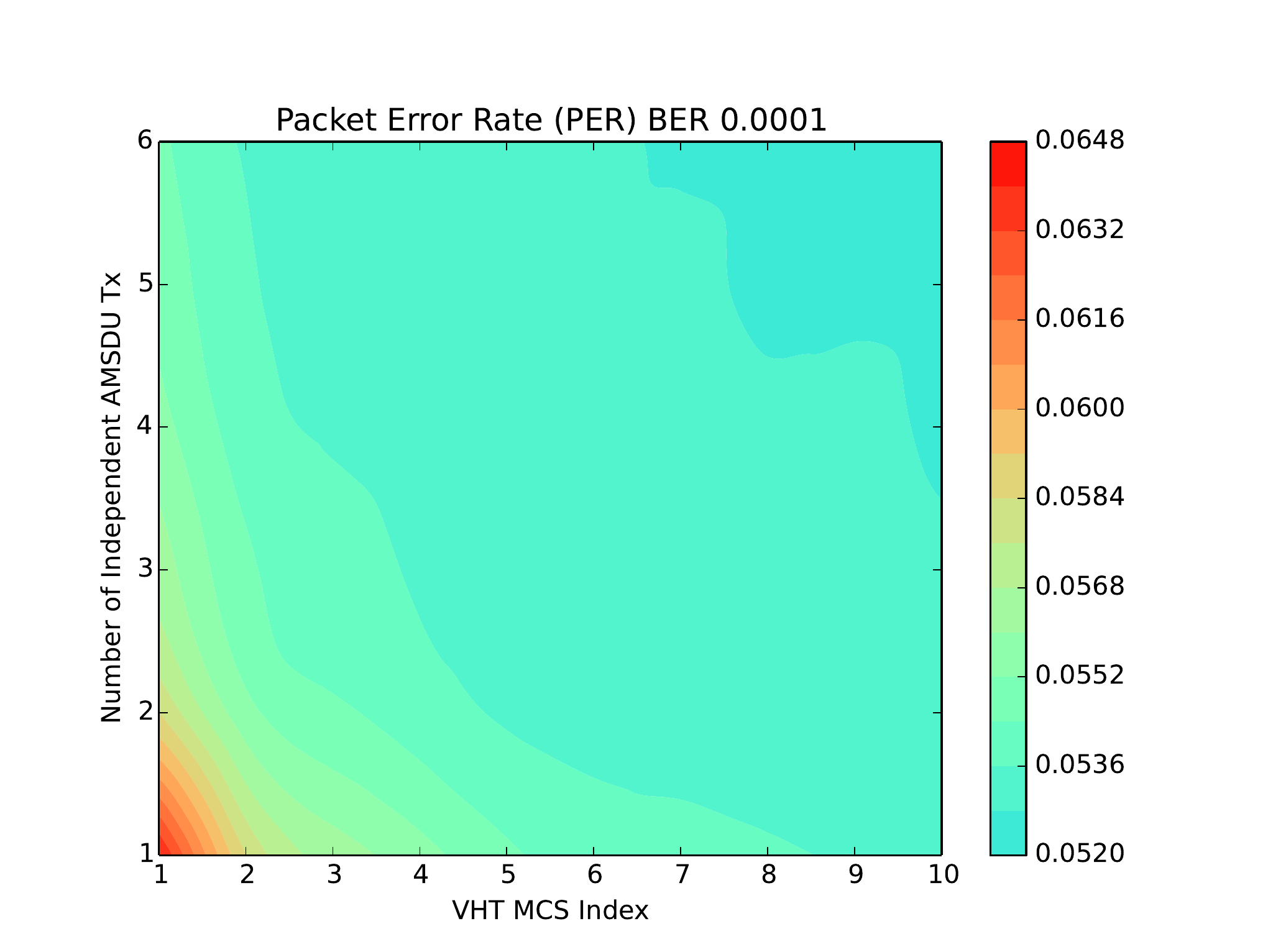,width=2.32in}
		\label{fig:ber0001}		
	\end{subfigure}
	\begin{subfigure}
		\centering
		\epsfig{figure=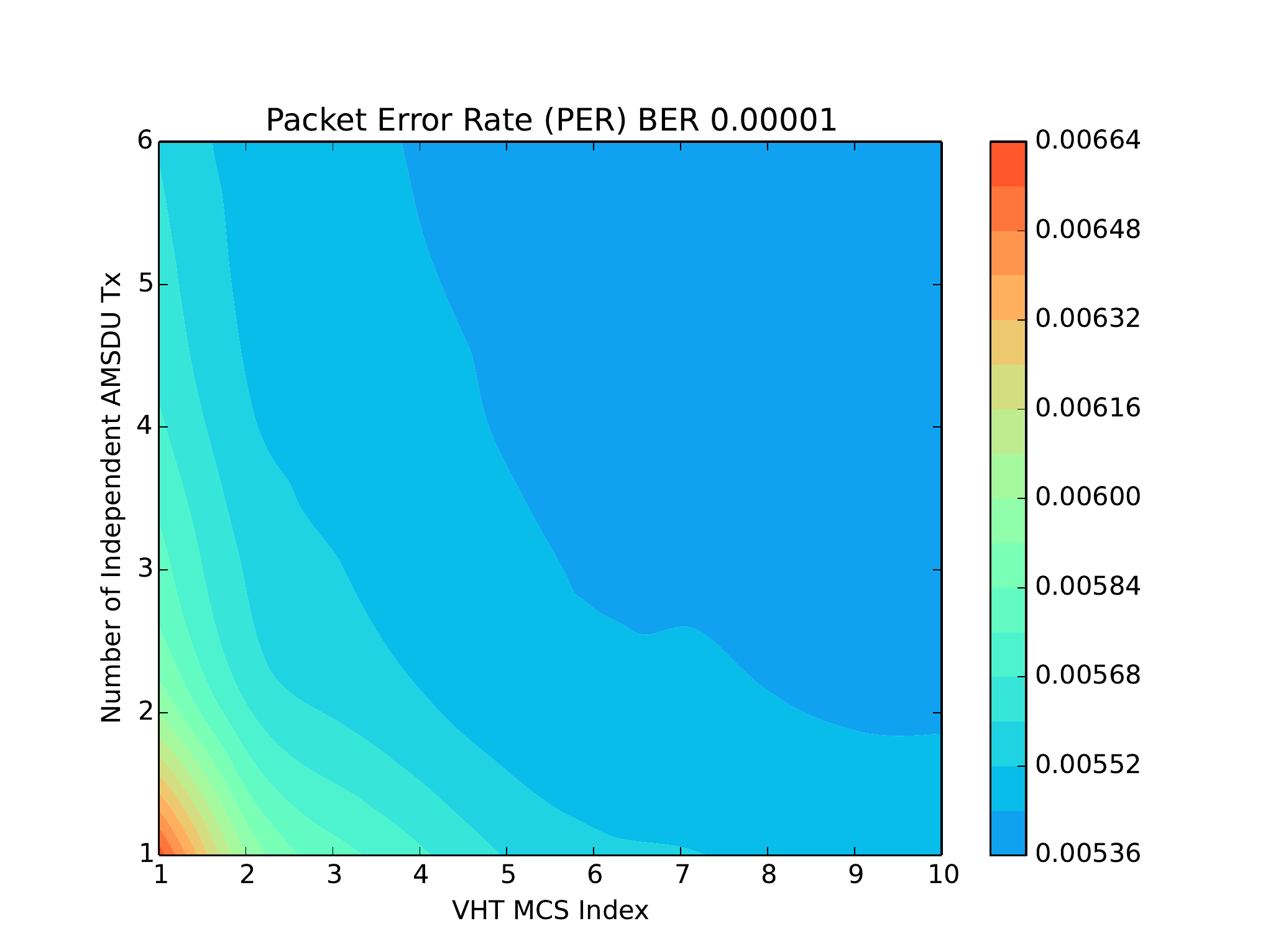,width=2.31in}
		\label{fig:ber0001}		
	\end{subfigure}
	\centering
	\caption{Variation in packet error rate (PE)R as a function of different AMSDU aggregation depths and VHT MCS rates. These PER values are plotted for different values of bit error rates (BER). }
	\label{fig:per}
\end{figure*}

\subsection{Packet Error Rate (PER) Performance}
PER performance experienced on a link determines the number of retries seen by a MAC protocol data unit (MPDU). In cases, where we have MSDU aggregation, the MPDU is also referred to as an aggregate MSDU or AMSDU. In this experiment, we vary the bit error rate (BER) and measure the effective PER achieved for different sizes of the AMSDU as a function of the MCS rates used. Results from the test are as shown in Figure~\ref{fig:per}. The X-axis is indicates the different MCS rates used for the transmission of the AMSDU, while the Y-axis represents the number of MSDU frames aggregated into the AMSDU. The higher the number of subframes in an AMSDU, the larger the packet size used on air.

A generic trend that can be seen from all the three subplots in Figure~\ref{fig:per} is that as the PER improves as we start using faster MCS rates. This is because as we approach faster frame transmissions speeds, the airtime used by the frame decreases and hence the collision probability also decreases.

As seen along the Y-axis of all of the plots in Figure~\ref{fig:per}, we also see that the packet error rate always drops when we are decreasing the number of MSDUs packed in an AMSDU. This happens because as the length of the frame decreases the collision probability of the frame also goes down.

\textbf{Inference: } The goal of this experiment was to check if we are able to achieve similar (desired) PER by dropping MCS rate for a given BER without having to change the size of the aggregate. As seen from the Figure~\ref{fig:per}, we observe that in most cases from any PER, to achieve a lower PER we could easily drop the MCS rate on the AMSDU without having to redo the AMSDU aggregation with fewer MSDUs. The potential downside of using such an approach is the use of a higher airtime for transmissions because of lower transmission rates. We evaluate the downside in the next set of experiments.

\begin{figure*}
	\centering
	\begin{subfigure}
		\centering
		\epsfig{figure=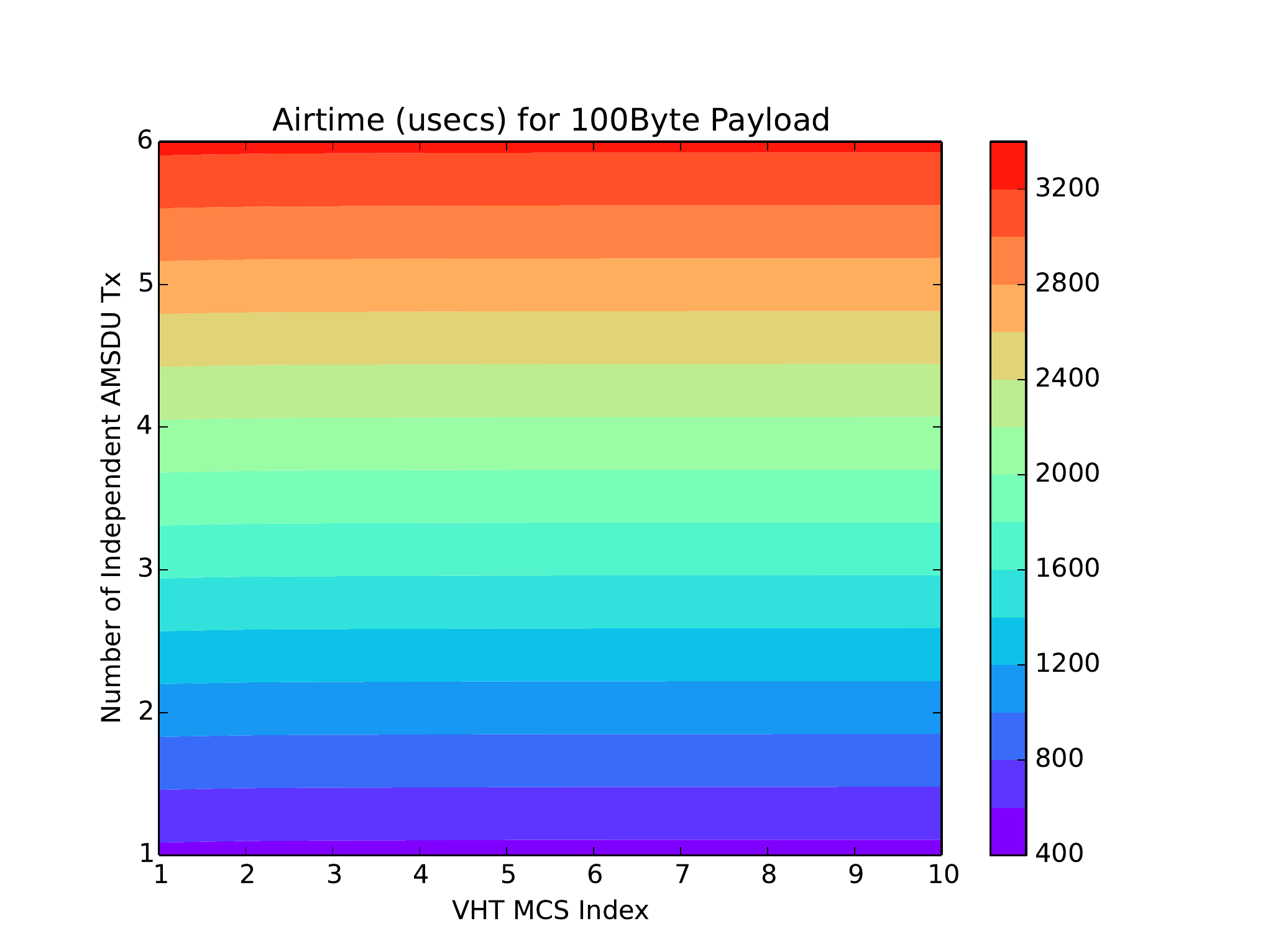,width=2.32in}
		\label{fig:hwv_rate_wts}
	\end{subfigure}
	\begin{subfigure}
		\centering
		\epsfig{figure=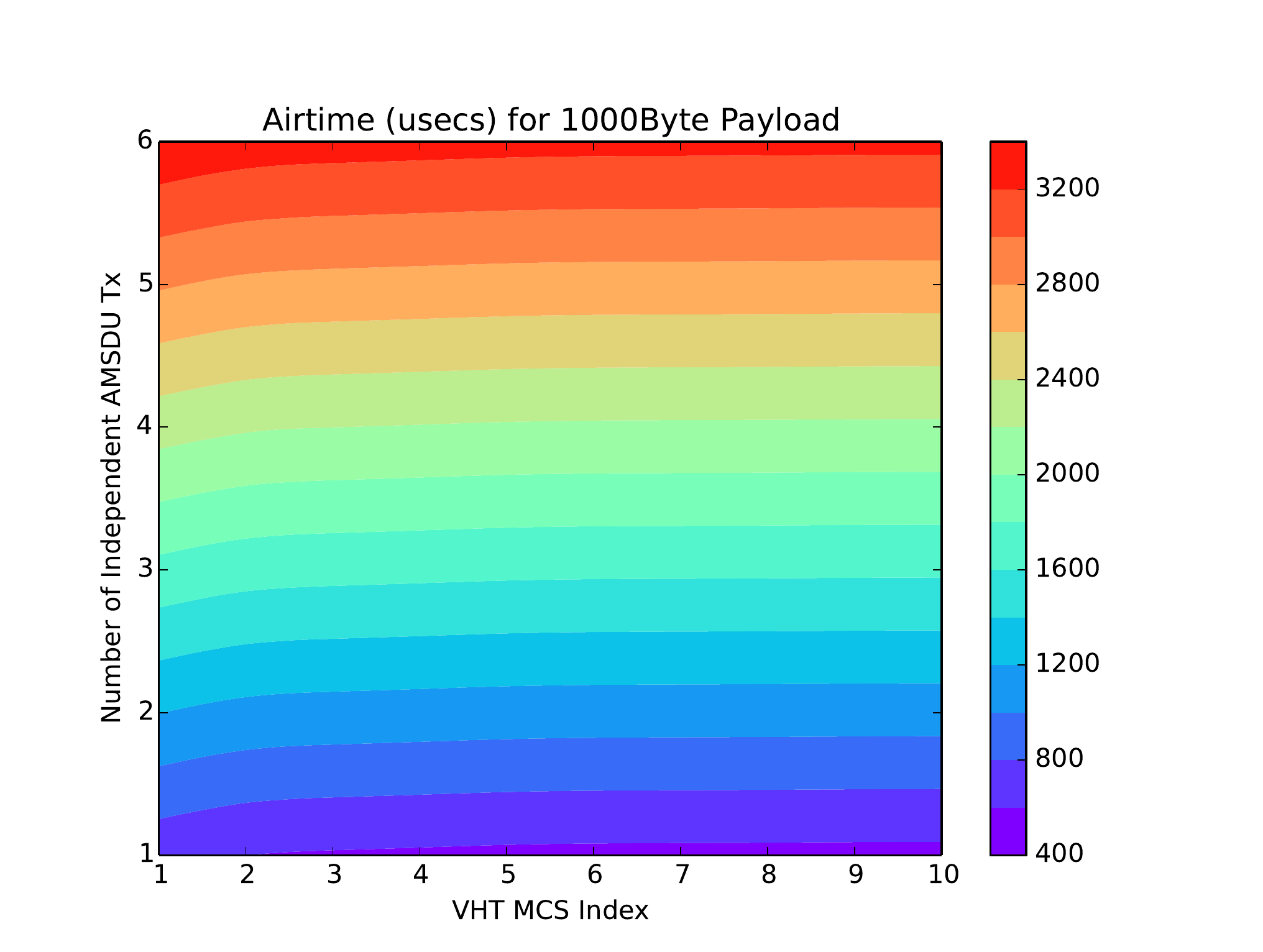,width=2.32in}
		\label{fig:hwv_rate_wts}		
	\end{subfigure}
	\begin{subfigure}
		\centering
		\epsfig{figure=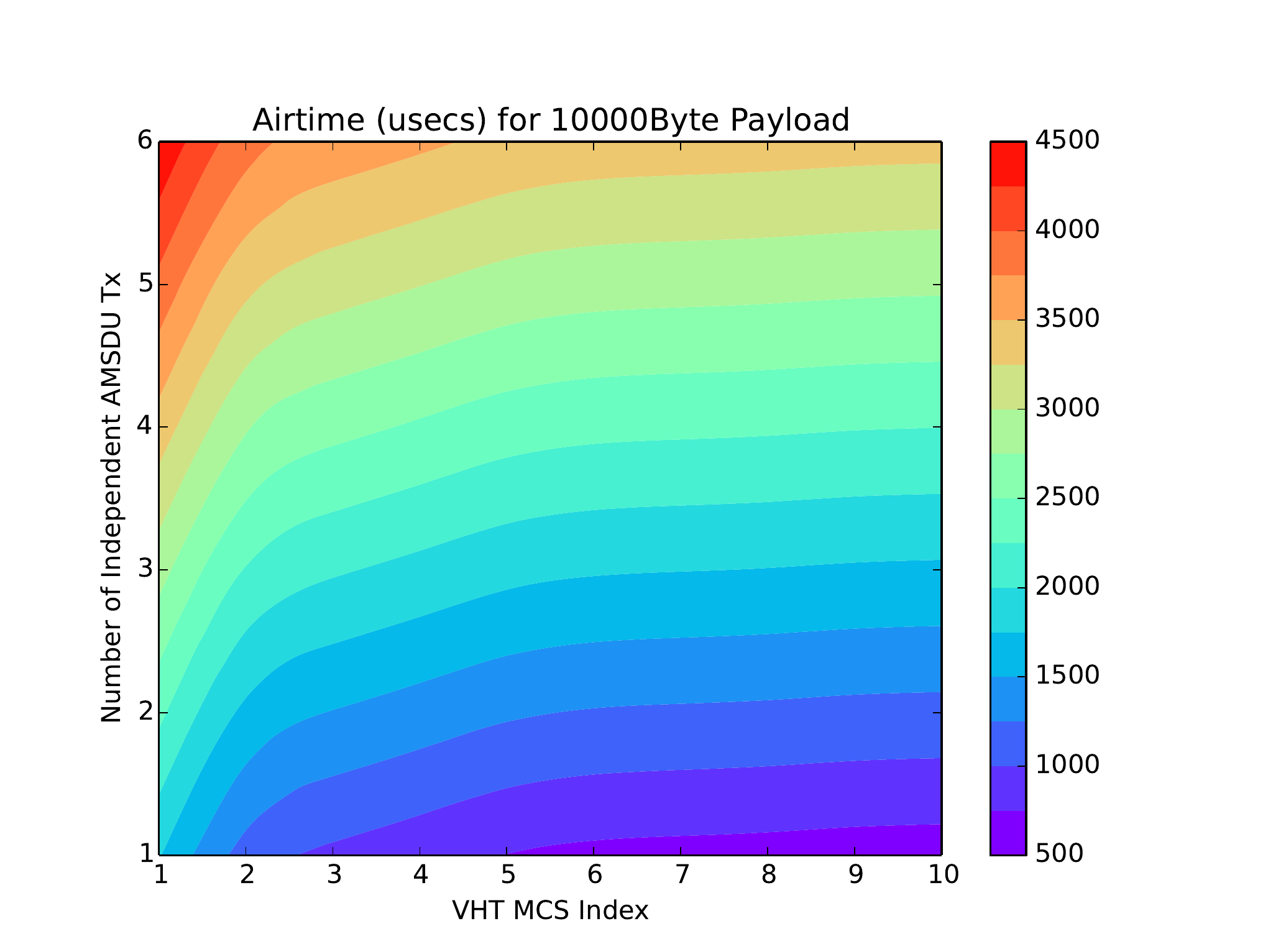,width=2.31in}
		\label{fig:PER_BER}		
	\end{subfigure}
	\centering
	\caption{Variation in airtime as a function of different MCS rates and aggregation depths. This is plotted for 3 different size of MSDUs: 100bytes, 1000bytes and 10Kbytes.}
	\label{fig:airtime}
\end{figure*}

\subsection{Airtime Performance}
Figure~\ref{fig:airtime} shows the amount of airtime required for transmitting AMSDUs at different aggregation depths as a function of the physical layer transmission MCS rates. These airtimes have been plotted for different MSDU payload sizes. The X-axis in these plots represent the MCS index used for the transmission. The Y-axis indicates the number of MSDUs packed in an AMSDU. The three plots are for different sizes of MSDU used in the AMSDU: 100bytes, 1Kbytes and 10Kbytes respectively.

We expect from the plots in Figure~\ref{fig:airtime} that as the MCS index drops the airtime increases. This effect is not at all seen for smaller packet sizes of the MSDU of 100bytes or 1Kbytes (subplot 1, 2) because at these sizes the airtime used is dominated by the MAC and Phy layer overheads of transmissions. These include but are not limited to MAC headers, PHY headers, and backoff. For larger MSDU sizes of 10Kbytes, we see some of this phenomenon. However, it is important to note that most of the packet sizes seen on the internet are close to 200bytes~\cite{internetpacketsize}, well within the 1Kbyte results shown here. It is also important to note that with newer WiFi standards supporting faster MCS rates~\cite{80211a},~\cite{80211n},~\cite{80211ax}, the impact on airtime seen at smaller MSDU sizes will decrease even further since those airtimes will tend to be dominated by the MAC PHY overheads.

From the plots in Figure~\ref{fig:airtime}, we also see that the airtime required for transmissions is higher for more MSDUs packed in an AMSDU. This is seen across all subplots. This is because as the frame size of transmission increases, the airtime required increases. We note that in the third subplot for MSDU size of 10Kbytes, the airtime for every aggregation depth plateaus, this is because of the limit of 5msec on the lenght of a PPDU transmission by the 802.11 standard.

\textbf{Inference: }As shown in the results from Figure~\ref{fig:airtime}, we observe that across almost all MSDU sizes that we considered, the airtime used for transmission does not deteriorate significantly with decreasing MCS rate.

\section{Conclusions and Future directions}
\label{sec:conc}
Results from our analysis show that (1) For most average packet sizes seen on the internet (less than 1KBytes), the airtime used for an AMSDU transmission is dominated by the depth of aggregation and not by the MCS used for transmissions. (2) Since the airtime does not increase significantly for dropping of rates, it is desired that we control PER by using rate control rather than adaptive frame aggregation at the AMSDU level. Discussions on the practicality of implementing adaptive AMSDU aggregation also show that re-framing AMSDUs in software have a significant performance impact in terms of re-queuing frames, encrypting and DMA transferring back to hardware.  Based on these simulations and arguments, we suggest controlling the AMSDU PER by dropping rates rather than implementing adaptive AMSDU frame aggregation. In the future, we plan to extend this study to encompass a wider set of MCS ranges and packet sizes.

\end{document}